\documentclass[preprint]{ptptex}

\title{
Spinor Representation of Conformal Group and Gravitational Model
}

\author{
Kohzo \textsc{Nishida}%
\footnote{E-mail: EZF01671@nifty.com} 
}

\inst{
Department of Physics, Kyoto Sangyo University, Kyoto 603-8555, Japan 
}

\abst{
We consider spinor representations of the conformal group. 
The spacetime is constructed by the 15-dimensional vectors in the adjoint representation of $SO(2,4)$. 
On the spacetime, we construct a gravitational model that is invariant under local transformation.
}

\begin{document}

\maketitle

Conformal transformations play a widespread role in gravity theories.\cite{1,2,3,4}
In this paper, we investigate the conformal transformations of spinors. 
The spinor, which describes spin-1/2 particles and antiparticles,\cite{5} contains a considerable amount of physical information, 
making it a promising target for investigation. 
In addition, we construct a gravitational model that is invariant under local transformation.

In general, an arbitrary $4\times 4$ transform matrix $U(x)$ that satisfies
\begin{equation}
 \label{eq:10}
  (\gamma_{0}U)^{\dagger}=(\gamma_{0}U)^{-1},
\end{equation}
has the generators:
\begin{equation}
 \label{eq:20}
  I,
  \,\,\,\,\,\,\,\,
  \gamma^i, 
  \,\,\,\,\,\,\,\,
  i\gamma_5,
  \,\,\,\,\,\,\,\,
  i\gamma^{ij}, 
  \,\,\,\,\,\,\,\,
  \gamma_5\gamma^{i}.
\end{equation}
$U(x)$ describes the conformal transformations. 
In fact, the following\cite{6,7}
\begin{eqnarray}
 \label{eq:30}
 M_{ij} \equiv \frac{i}{2}\gamma_{ij}, 
 \,\,\,\,\,\,
 P_i \equiv \frac{1}{2}(\gamma_i +\gamma_5 \gamma_i), \nonumber \\
 D \equiv \frac{i}{2}\gamma_5, 
 \,\,\,\,\,\,
 K_i \equiv \frac{1}{2}(\gamma_i - \gamma_5 \gamma_i),
\end{eqnarray}
satisfies the conformal group
\begin{eqnarray}
 \label{eq:40}
&&[D, K_i] = -iK_i, \,\,\,\,\,\,[D, P_i] = iP_i \nonumber \\
&&[ K_i, P_j] = 2i\eta_{ij}D - 2iM_{ij},  \,\,\,\,\,\,[K_i, M_{jk}] = i(\eta_{ij}K_k - \eta_{ik}K_j), \nonumber \\
&&[P_i, M_{jk}] = i(\eta_{ij} P_k - \eta_{ik}P_j), \nonumber \\
&&[M_{ij}, M_{kl}] = i(\eta_{jk}M_{il} + \eta_{il}M_{jk} - \eta_{ij}M_{jl} - \eta_{jl}M_{ik}).
\end{eqnarray}
Other commutators vanish. Here, we have constructed a flat tangent space at every point on the four-dimensional manifold, 
and we indicate the vectors in the four-dimensional tangent space by subscript Roman letters $i$, $j$, $k$, etc. 
Greek letters, such as $\mu$ and $\nu$, are used to label four-dimensional spacetime vectors. 
$U(x)$ contains within it local Lorentz transformations. Let a spinor $\psi$ transform as
\begin{equation}
 \label{eq:50}
  \psi \rightarrow \psi' = U\psi,
\end{equation}
then (\ref{eq:30}) is spinor representations of the conformal group.
If we rewrite (\ref{eq:30}) as 
\begin{equation}
 \label{eq:60}
  J_{ij} \equiv M_{ij}, \,\,\,\,\,\,J_{i4} \equiv \frac{1}{2}(K_i - P_i), \,\,\,\,\,\,J_{i5} \equiv \frac{1}{2}(K_i + P_i),\,\,\,\,\,\,J_{54} \equiv D.
\end{equation}
$J_{pq}$ generates the algebra of $SO(2,4)$:
\begin{equation}
 \label{eq:70}
[J_{pq}, J_{rs}] = i(\eta_{qr}J_{ps} + \eta_{ps}J_{qr} - \eta_{pq}J_{qs} - \eta_{qs}J_{pr}), 
\end{equation}
where $\eta_{pq} = (1,-1,-1,-1,1)$ and $p,q,r,s = 0,1,\cdots 5$.

$U(x)$ can be also written as
\begin{equation}
 \label{eq:80}
  U(x) = e^{-i\varepsilon_a (x) \Gamma^a},
\end{equation}
where 
\begin{eqnarray}
 \label{eq:90}
\lefteqn{2\Gamma^a} \nonumber \\
&\equiv& 
(
\gamma^0, \gamma^1, \gamma^2, \gamma^3, i\gamma_5, 
i\gamma^{01}, i\gamma^{02}, i\gamma^{03}, i\gamma^{12}, i\gamma^{13}, i\gamma^{23}, 
\gamma_5 \gamma^0, \gamma_5 \gamma^1, \gamma_5 \gamma^2, \gamma_5 \gamma^3
) .
\end{eqnarray}
We have projected out the unit matrix $I$.
These gamma matrices satisfy
\begin{eqnarray}
 \label{eq:100}
&& {\rm Tr}(\Gamma^a \Gamma^b) = 0 \,\,\,\,a \neq b, \nonumber \\
&& {\rm Tr}(\Gamma^a \Gamma^a) = 1\,\,\,\,a=0,8,9,10,12,13,14, \nonumber \\
&& {\rm Tr}(\Gamma^a \Gamma^a) = -1\,\,\,\,a=1,2,3,4,5,6,7,11.
\end{eqnarray}
Now, let us suppose that the spacetime transforms as the 15-dimensional vectors in the adjoint representation of $SO(2,4)$, 
not the six-dimensional vectors in the fundamental representation, 
that is, the 15-dimensional coordinates $x_a(a=0,1,\cdots 14)$ transform as 
\begin{equation}
 \label{eq:110}
X\rightarrow X' = U^{-1} X U,\,\,\,\,\,\,X \equiv x_a \Gamma^a.
\end{equation}
Note that $\gamma_{0}U$ with the generators  (\ref{eq:20}) is the transform matrix of $U(4)$. 
The space coordinates are rotated in the adjoint representation of $SU(2)$. 
Therefore, it is not unnatural that the spacetime transforms in the adjoint representation. 
An invariant can be constructed:
\begin{eqnarray}
 \label{eq:120}
{\rm Tr}(XX) &=& x^2_0 -x^2_1 -x^2_2 -x^2_3 -x^2_4 -x^2_5 -x^2_6 -x^2_7 
\nonumber \\
&& +x^2_8 +x^2_9 +x^2_{10}-x^2_{11} +x^2_{12} +x^2_{13} +x^2_{14}. 
\end{eqnarray}
Then, the metric of the tangent space is $\eta_{ab} = (1,-1,-1,-1,-1,-1,-1,-1,1,1,1,-1,1,1,1)$.
Let us define a 15-dimensional tetrad 
\begin{equation}
 \label{eq:130}
e^{a\alpha} e_{a}\,^{\beta} = g^{\alpha\beta},
\,\,\,\,\,\,
e^{a}\,_{\alpha} e^{b\alpha} = \eta^{ab} ,
\end{equation}
where indices $a,b,c$ are 15-dimensional vectors in flat space and $\alpha, \beta, \gamma$ are 15-dimensional vectors in curved space. 
The 15-dimensional tetrad transforms linearly as 
\begin{equation}
 \label{eq:140}
e_{a}\,^{\alpha}(x) \Gamma^a \rightarrow  U^{-1}(x) \Gamma^\alpha(x) U(x)= e_{a}\,^{\alpha}(x)U^{-1}(x) \Gamma^a U(x) =e'_{a}\,^{\alpha}(x') \Gamma^a.
\end{equation}

Next, we propose a gauge model on this spacetime. 
In our previous paper\cite{8}, we have considered the four- and five-dimensional gauge model. 
The covariant derivative is given by
\begin{equation}
 \label{eq:150}
  D_{\alpha} \equiv \partial_\alpha - igA_{\alpha},\,\,\,\,\,\,
A_\alpha \equiv A_{a\alpha}\Gamma^a,
\end{equation}
where 
we introduce gauge fields $A_{\alpha}$, each transforming as 
\begin{equation}
 \label{eq:160}
  A_{\alpha} \rightarrow A'_\alpha = U A_\alpha U^{-1} + \frac{1}{ig}U \partial_\alpha U^{-1}.
\end{equation}
The covariant derivative transforms as
\begin{equation}
 \label{eq:170}
  D_{\alpha} \rightarrow D'_\alpha = U D_\alpha U^{-1}.
\end{equation}
In general, the covariant derivative of the tetrad should be zero\cite{9}:
\begin{eqnarray}
 \label{eq:180}
  \partial_{\mu}e_{i\nu}+g\omega_{ij,\mu}e^{j}\,_{\nu}-\Gamma^{\lambda}\,_{\nu\mu}e_{i\lambda}=0,
\end{eqnarray}
where $\Gamma^{\lambda}\,_{\nu\mu}$ is the Christoffel symbol and 
$\omega_{ij,\mu}$ is the spin connection.
(\ref{eq:180}) can be rewritten as
\begin{equation}
 \label{eq:190}
 [D_\mu, \gamma_\nu] = \Gamma^{\lambda}\,_{\nu\mu}\gamma_{\lambda},
\end{equation}
where
\begin{equation}
 \label{eq:200}
D_\mu \equiv \partial_\mu + g\omega_\mu,
\,\,\,\,\,\,
\omega_\mu \equiv \frac{i}{4} \omega_{ij, \mu} \gamma^{ij}.
\end{equation}
In direct analogy, we require
\begin{equation}
 \label{eq:210}
  [D_\alpha, \Gamma_\beta] = \Gamma^{\gamma}\,_{\beta\alpha}\Gamma_{\gamma}.
\end{equation}
Using this, we can calculate $[ [D_\alpha, D_\beta], \Gamma_\gamma]$ as
\begin{eqnarray}
 \label{eq:220}
  [ [D_\alpha, D_\beta], \Gamma_\gamma]
 &=& [ D_\alpha, [D_\beta, \Gamma_\gamma]] - [ D_\beta, [D_\alpha, \Gamma_\gamma]] \nonumber \\
 &=& \{ \partial_\alpha \Gamma^\delta\,_{\gamma \beta} + \Gamma^\delta\,_{\epsilon \alpha} \Gamma^\epsilon\,_{\gamma \beta}
 -(\alpha \leftrightarrow \beta) \} \Gamma_\delta \nonumber \\
 &=& R^{(15)\delta}\,_{\gamma, \alpha \beta} \Gamma_\delta,
\end{eqnarray}
where $R^{(15)\delta}\,_{\gamma, \alpha \beta} 
\equiv \partial_\alpha \Gamma^\delta\,_{\gamma \beta} -\partial_\beta \Gamma^\delta\,_{\gamma \alpha} 
+ \Gamma^\delta\,_{\epsilon \alpha} \Gamma^\epsilon\,_{\gamma \beta} - \Gamma^\delta\,_{\epsilon \beta} \Gamma^\epsilon\,_{\gamma \alpha}$
 is the 15-dimensional Riemann curvature tensor.
We multiply both sides of (\ref{eq:220}) by $\Gamma_\zeta$ and consider the trace to obtain
\begin{equation}
 \label{eq:230}
R^{(15)}_{\gamma \delta, \alpha \beta} = -2{\rm Tr}(\Gamma_{\gamma\delta}[D_\alpha, D_\beta]),
\,\,\,\,\,\,
\Gamma^{\alpha\beta} \equiv \frac{1}{2}[\Gamma^\alpha, \Gamma^\beta].
\end{equation}
Therefore, the 15-dimensional Riemann curvature tensor is invariant under the transformation of $U(x)$.

Let us now try to form an action with this formalism.
We find that the following Lagrangian
\begin{equation}
 \label{eq:240}
  {\cal L}=  e  \left\{ \bar{\psi}(2i\Gamma_\alpha D^{\alpha} - m) \psi +\frac{1}{16\pi G^{(15)}}R^{(15)} \right\}
\end{equation}
is invariant under the transformation of $U(x)$, 
where $R^{(15)}$ is the 15-dimensional Ricci curvature and $G^{(15)}$ is the 15-dimensional constant of gravitation.
The Dirac's equation is 
\begin{equation}
 \label{eq:250}
(2i\Gamma^a \partial_a -m )\psi =0.
\end{equation}
On multiplying (\ref{eq:250}) by $2\Gamma^b \partial_b$ from the left, we obtain the high-dimensional Klein--Cordon equation:
\begin{eqnarray}
 \label{eq:260}
&&(\partial_i \partial^i - \partial^{(5)}\partial^{(5)} +\partial_{ij}\partial^{ij}-\partial^{(5)}_i \partial^{(5)i} -i\varepsilon^{ijkl} \gamma_5 \partial_{ij} \partial_{kl} \nonumber \\
&& +2i\gamma^{ij}\partial^{(5)}\partial_{ij} -2i\varepsilon^{ijkl} \gamma_{kl} \partial_i \partial^{(5)}_j -2i\varepsilon^{ijkl} \gamma_{kl} \partial^{(5)} \partial_{ij}\nonumber \\
&&-2\varepsilon^{ijkl}\gamma_5 \gamma_l \partial_i \partial_{jk}  + 2i\varepsilon^{ijkl}\gamma_k \partial_{ij}\partial^{(5)}_k)\psi + m^2 \psi = 0,
\end{eqnarray}
where 
the summation of $\partial_{ij}$ are limited to $i<j$ and 
we represent the 15-dimensional coordinates corresponding to (\ref{eq:90}) as 
\begin{equation}
 \label{eq:270}
x^a = (
x^0, x^1, x^2, x^3, x^{(5)}, 
x^{01}, x^{02}, x^{03}, x^{12}, x^{13}, x^{23}, 
x^{(5)0}, x^{(5)1}, x^{(5)2}, x^{(5)3}
).
\end{equation}

Our model still cannot explain why the conformal symmetry is broken. 
However, it is interesting that the extra dimension and the four-dimensions have a different spatial structures compared with common high-dimensional models. 
Through conformal symmetry breaking, our model would undergo the reduction from 15 down to four dimensions, and the gauge-fixed tetrad must be
\begin{equation}
 \label{eq:3100}
e^{i\mu} e_{i}\,^{\nu} = g^{\mu\nu}
\,\,\,\,\,\,
e^{i}\,_{\mu} e^{j\mu} = \eta^{ij} ,
\,\,\,\,\,\,
i,j,\mu,\nu=0,2,3,4.
\end{equation}
Note that the following tetrad also constructs four-dimensional spacetime:
\begin{equation}
 \label{eq:3200}
e^{i\mu} e_{i}\,^{\nu} = g^{\mu\nu}
\,\,\,\,\,\,
e^{i}\,_{\mu} e^{j\mu} = \eta^{ij} ,
\,\,\,\,\,\,
i,j=11,12,13,14,
\,\,\,\,\,\,
\mu,\nu=0,2,3,4.
\end{equation}
This would seem to suggest that there is a new internal degree of freedom in our spacetime.

\end{document}